\documentclass[%
reprint,
superscriptaddress,
nobibnotes,
 amsmath,amssymb,
prb,
]{revtex4-1}

\usepackage{graphicx}
\usepackage{dcolumn}% Align table columns on decimal point
\usepackage{bm}% bold math
\usepackage{multirow}
\usepackage{mathptmx}
\usepackage{siunitx}
\usepackage{color}
	
\usepackage{lineno}
\graphicspath{{./img/}}

\newcommand{\beq}[1]{\begin{equation}\label{#1}}
\newcommand{\eep}{\;.\end{equation}}
\newcommand{\eec}{\;,\end{equation}}
\newcommand{\eeq}{\end{equation}}

\newcommand{\lb}{\left(}
\newcommand{\rb}{\right)}

\renewcommand{\a}{\alpha}

\newcommand{\eo}{{\epsilon}_0}
\newcommand{\ep}{\epsilon}

\newcommand{\Ep}{\mathcal{E}}

\renewcommand{\th}{\theta}

\newcommand*\dd{\mathop{}\!\mathrm{d}} %differential d
\newcommand{\grad}{\nabla} % Grad

\newcommand{\dmin}{d_{\text{min}}}

\newcommand*\chem[1]{\ensuremath{\mathrm{#1}}} % for chemical symbols

\newcommand{\V}{\mathcal{V}} % total energy as an integral

\newcommand{\Asc}{A_{\text{sc}}} % area of supercell
\newcommand{\Ct}{\mathcal{C}_3}
\newcommand{\Cs}{\mathcal{C}_6}

\DeclareMathAlphabet{\mathcal}{OMS}{cmsy}{m}{n} % Changes font for mathcal but leaves the rest of the math fonts in Times.

\begin{document}

%%% TITLE, AUTHORS, ABSTRACT%%%

\title{On electrically tunable stacking domains and ferroelectricity in moir\'e superlattices}

\author{Daniel Bennett}
\email{db729@cam.ac.uk}
 \affiliation{Theory of Condensed Matter, Cavendish Laboratory, Department of Physics, J J Thomson Avenue, Cambridge CB3 0HE, United Kingdom}
 
\author{Benjamin Remez}
\affiliation{Theory of Condensed Matter, Cavendish Laboratory, Department of Physics, J J Thomson Avenue, Cambridge CB3 0HE, United Kingdom}

\date{\today}

\begin{abstract}
\textbf{\abstractname.} It is well known that stacking domains form in moir\'e superlattices due to the competition between the interlayer van der Waals forces and intralayer elastic forces, which can be recognized as polar domains due to the local spontaneous polarization in bilayers without centrosymmetry. We propose a theoretical model which captures the effect of an applied electric field on the domain structure. The coupling between the spontaneous polarization and field leads to uneven relaxation of the domains, and a net polarization in the superlattice at nonzero fields, which is sensitive to the moir\'e period. We show that the dielectric response to the field reduces the stacking energy and leads to softer domains in all bilayers. We then discuss the recent observations of ferroelectricity in the context of our model.
\end{abstract}

\pacs{Valid PACS appear here}% PACS, the Physics and Astronomy

\maketitle

%%% MAIN TEXT %%%

\section*{Introduction}

Twistronics, the study of layered systems with a relative twist angle or lattice mismatch between the layers, resulting in moir\'e superlattices, is one of the most exciting new topics in condensed matter physics. It was predicted about a decade ago that introducing a small relative twist in a layered system such as bilayer graphene could lead to flat electronic bands, and strongly correlated behavior\cite{morell2010flat,bistritzer2011moire}. Moir\'e superlattices have since been shown to exhibit superconductivity\cite{cao2018unconventional,yankowitz2019tuning}, metal-insulator transitions\cite{cao2018correlated}, as well as magnetic\cite{sharpe2019emergent}, topological\cite{yin2016direct,rickhaus2018transport,huang2018topologically,sunku2018photonic} and excitonic\cite{yu2017moire,seyler2019signatures} behavior, facilitated by the tuning of the twist angle or lattice mismatch. Recently, ferroelectricity was observed in bilayer graphene\cite{zheng2020unconventional} and hexagonal boron nitride (hBN)\cite{stern2020interfacial}, which is highly unusual, because the constituent materials are non-polar, and bilayer graphene is normally metallic. The ferroelectricity was found to be sensitive to the twist angle and lattice mismatch, with some samples exhibiting no hysteresis and some exhibiting strong hysteresis. The ferroelectricity is clearly very unconventional, and the physical mechanism is currently not well understood.

\begin{figure}[h]
\includegraphics[width=\columnwidth]{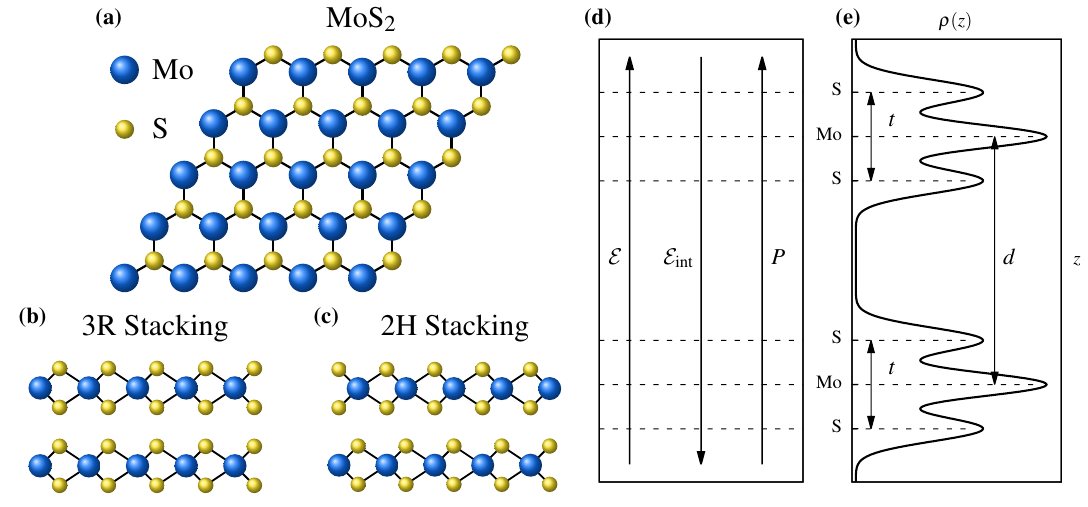}
\caption{Atomic and electrostatic sketches of bilayer \chem{Mos_2}. \textbf{(a)}: Atomic structure of a \chem{MoS_2} monolayer. The two stacking configurations are shown below: \textbf{(b)}: alignment of the layers (3R stacking) and \textbf{(c)}: the two layers mirrored with respect to one another (2H stacking or a twist of $180^{\circ}$). \textbf{(d)}: Sketch of the electrostatics of bilayer \chem{MoS_2}. The dashed lines indicate the vertical positions of the atoms, and the vectors show the applied field $\Ep$ and resulting polarization $P$ and internal field $\Ep_{\text{int}}$. \textbf{(e)}: Sketch of the charge density of bilayer \chem{MoS_2} along the out-of-plane direction, averaged in the in-plane directions.}
\label{fig:Fig1}
\end{figure}

Structural phenomena in moir\'e superlattices are generally well understood. It is known that the interlayer separation ripples in space due to the local misalignment of the atoms, which can influence physical properties \cite{guinea2008gauge,san2014spontaneous}. Additionally, lattice relaxation occurs due to the competition between the in-plane strains and out-of-plane van der Waals interactions, leading to stacking domains\cite{jung2015origin,nam2017lattice,zhang2018structural,carr2018relaxation,
lebedeva2019commensurate,lebedeva2019energetics,lebedeva2020two}. The elastic energy depends on the twist angle and lattice mismatch quadratically, meaning the domains can be tuned. The domain structures have been shown to have a large influence on the properties of the system\cite{alden2013strain,woods2014commensurate,yankowitz2016pressure,
jung2015origin,nam2017lattice,carr2017twistronics,zhang2018structural}, leading to the opening of band gaps and enhanced Fermi velocity, for example. Polar effects have been given less consideration because the typical materials use to fabricate moir\'e superlattices, graphene, hBN and transition metal dichalcogenides (TMDs) such as \chem{MoS_2} (see Figs.~\ref{fig:Fig1} (a)-(c)), are non-polar. 

There are two main mechanisms by which polar phenomena can manifest in moir\'e systems. The first is a local spontaneous out-of-plane polarization\cite{li2017binary}, which occurs in bilayers without centrosymmetry and averages to zero over the moir\'e period. The second is couplings between strain and polarization, namely piezoelectricity\cite{mcgilly2020visualization,enaldiev2021piezoelectric,ferreira2021weak} and flexoelectricity\cite{mcgilly2020visualization,mashkevich1957electrical,tolpygo1962investigation,indenbom1981flexoelectric,kogan1964piezoelectric,tagantsev1985theory,zubko2013flexoelectric}. The strain gradient is largest across the domain walls, and via flexoelectricity, they have an inherent polarization. The flexoelectric response in 2D materials can be estimated by measuring the potential drop across the wall of a nanotube in the large radius limit\cite{PhysRevB.90.201112,mcgilly2020visualization,artyukhov2020flexoelectricity,bennett2021flexoelectric}, and it has been estimated that the flexoelectric coefficients in bilayer graphene is similar in magnitude to the clamped-ion flexoelectric response in oxide perovskites\cite{mcgilly2020visualization,PhysRevB.90.201112,dreyer2018current}. The flexoelectric polarization is localized within the relatively narrow domain walls, however.

If we identify the stacking domains as polar domains via the two aforementioned mechanisms, then the stacking domains may serve as the basis for understanding polar phenomena in moir\'e materials. Thus, in order to understand the observed ferroelectricity, it is essential to understand how the stacking domains respond to an electric field. It is known that the domain structures in moir\'e materials can lead to interesting effects such as the opening of band gaps, and topologically protected states or channels when an electric field is applied\cite{vaezi2013topological,yin2016direct,rickhaus2018transport,huang2018topologically,efimkin2018helical}. To our knowledge, the influence of an applied electric field on the domains themselves has not been considered. It is known that an electric field can modify the interlayer separation and lead to a breakdown of TMD bilayers, for example\cite{santos2013electrically,li2018asymmetric}. The stacking domains are a result of lattice relaxation, which describes the delicate competition between the interlayer interactions and the intralayer elasticity. Since the interlayer interactions are sensitive to an applied field, it is reasonable to expect that the field would change the delicate balance and affect the resulting domain structure.

\section*{Results}

In this paper, we introduce a model of lattice relaxation in a moir\'e superlattice which includes the effect of an applied field on the bilayer. The total energy is an integral of the energy density over a moir\'e supercell
\beq{eq:V_tot_main}
V_{\text{tot}} = \frac{1}{\Asc}\int_{\Asc} \V_{\text{tot}}(\mathbf{r})\dd\mathbf{r}
\eec
where $\Asc$ is the area of the supercell. For a bilayer system Eq.~\eqref{eq:V_tot_main} is a discrete sum over atomic sites, but generalizes to a continuum field theory when the moir\'e period is much larger than the lattice constants of the monolayers. 

We can model moir\'e superlattices at different levels of theory depending on the contributions we include in Eq.~\eqref{eq:V_tot_main}. The stacking energy $\V_{\text{stack}}$ captures the weak van der Waals interactions between the layers in terms of the layer separation $d$. The elastic energy $\V_{\text{elastic}}$ allows for in-plane displacements of the atoms in the layers. Together, the stacking and elastic energies provide a good description of the atomic structure in moir\'e superlattices, namely lattice relaxation and the formation of stacking domains. Having obtained a realistic description of the structure, one could proceed to obtain the electronic bands from tight-binding theory.

In order to consider the effect of an electric field on the atomic structure, we also include the electrostatic energy induced by an electric field $\Ep$ perpendicular to the bilayer, see Fig.~\ref{fig:Fig1} (d). The total energy density is then
\beq{eq:V_tot_all}\resizebox{0.85\columnwidth}{!}{$
\begin{gathered}
\V_{\text{tot}} = \V_{\text{elastic}} + \V_{\text{stack}} + \V_{\text{elec}}\\[10pt]
\V_{\text{elastic}} = C_{ijkl}\ep_{ij}\ep_{kl}\\
\V_{\text{stack}} = \left|\V_0(\mathbf{r})\right|\left[\lb\frac{d_0(\mathbf{r})}{d}\rb^{12}-2\lb\frac{d_0(\mathbf{r})}{d}\rb^6 \right]\\
\V_{\text{elec}} = - \Ep p_0(\mathbf{r}) - \frac{1}{2}\eo\lb \a_0(\mathbf{r}) + \a_1(\mathbf{r})\lb \frac{d}{d_0(\mathbf{r})}-1\rb\rb\Ep^2
\end{gathered}
$}
\eep
Each contribution is derived and discussed in detail in Section I of the supplementary information. In the elastic energy, summation is assumed, $C$ is the linear elasticity tensor and ${\ep_{ij} = \frac{1}{2} \lb \partial_i U_j(\mathbf{r}) + \partial_j U_i(\mathbf{r})\rb}$ is the strain tensor, written in terms of a relative in-plane displacement $\mathbf{U}$.

The stacking energy can be included in a number of ways. The simplest is to use the cohesive energy as a function of space, $\V_0(\mathbf{r})$, assuming that at each point in the supercell the layer separation takes the value that minimizes the local stacking energy: $d(\mathbf{r}) = d_0(\mathbf{r})$. Other studies have allowed the layer separation to vary by performing a harmonic expansion about the equilibrium layer separation\cite{guinea2008gauge}. When considering the effect of an applied field, it is necessary to include the full van der Waals potential because some phenomena cannot be captured at the harmonic level, such as the breakdown of the bilayer for stronger fields \cite{santos2013electrically}. A detailed study of the stacking energy of bilayer \chem{MoS_2} in the presence of an electric field, both theoretically and verified by first-principles calculations, is provided in Appendices A and B, respectively.

The first term in $\V_{\text{elec}}$ is the coupling between the electric field and the out-of-plane spontaneous dipole moment of the bilayer\cite{li2017binary}. Bilayer systems without centrosymmetry, such as 3R \chem{MoS_2} (Fig.~\ref{fig:Fig1} (b)), have a local dipole moment throughout the supercell which averages to zero, whereas systems with centrosymmetry have no local dipole moment anywhere in the supercell, such as 2H \chem{MoS_2} (Fig.~\ref{fig:Fig1} (c)).  

The second term describes the dielectric response of the bilayer to the electric field, where $\a_0$ and $\a_1$ are the first two coefficients in the expansion of the polarizability $\a$ about the equilibrium layer separation. A bilayer system cannot simply be treated as a pair of capacitor plates; due to the overlap of states in the vacuum region between the layers, it is more appropriate to treat the system as a single slab with a nonuniform charge density, see Fig.~\ref{fig:Fig1} (e). Thus, changing the layer separation will affect the polarizability of the system, so we perform a Taylor expansion in $d$. In Section II of the supplementary information we show with first-principles calculations that the polarizability is linear in $d$. In addition, the polarizability will vary throughout the superlattice due to the different local stacking configurations and equilibrium layer separations. The dielectric response occurs in all layered systems, irrespective of symmetry.

The typical lattice relaxation procedure is as follows: the local energy densities in Eq.~\eqref{eq:V_tot_all} are parameterized using first-principles calculations. Practically, this is done using the mapping between real space and `configuration space' \cite{carr2018relaxation}, where all of the local stacking configurations in real space are condensed into a single unit cell of relative translations between the layers (see Section II of the supplementary information). Configuration space can be traversed using first-principles calculations with only a single primitive cell (6 atoms for bilayer \chem{MoS_2}), and translating one layer over the other. Quantities such as $\V_0$, $d_0$, etc., can be parameterized in configuration space, and Eq.~\eqref{eq:V_tot_main} can then be minimized with respect to the layer separation $d$ and in-plane displacements $U$ in order to obtain the relaxed structure. 

\begin{figure}[h]
\centering
\hspace*{-0.25cm}
\includegraphics[width=\columnwidth]{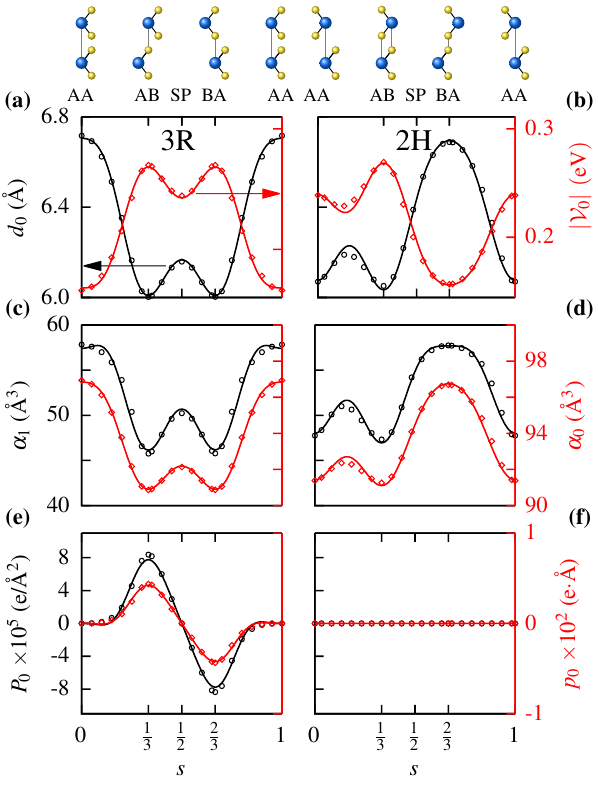}
\caption{
Parameterization of Eq.~\eqref{eq:V_tot_all} along the configuration space diagonal for 3R and 2H stacked bilayer \chem{MoS_2}. \textbf{(a, c, e)}: 3R-stacked bilayer \chem{MoS_2}: 
\textbf{(a)} $d_0$ (black) and $\left|\V_0\right|$ (red);
\textbf{(c)} $\a_1$ (black) and $\a_0$ (red);
\textbf{(e)} $P_0$ (black) and $p_0$ (red).
The hollow markers are results from first-principles calculations, and the solid curves show the corresponding Fourier interpolations. Black refers to the leftmost vertical axis and red to the rightmost vertical axis.
The configurations AA $\lb s = 0 , 1\rb$, AB $\lb s = \tfrac{1}{3}\rb$, SP $\lb s = \tfrac{1}{2}\rb$ and BA $\lb s = \tfrac{2}{3}\rb$ are marked and sketched above the plots. \textbf{(b, d, f)}: The same quantities, computed for 2H stacking. }
\label{fig:Fig2}
\end{figure}

The parameterization of $\V_0$, $d_0$, $\a_0$, $\a_1$ and $P_0$ for 3R and 2H \chem{MoS_2} was done using {\sc siesta}\cite{siesta} and is shown in Fig.~\ref{fig:Fig2}. Starting from the metal over metal configuration (AA), one layer was fixed and the other was translated along primitive cell diagonal in small increments, and the aforementioned quantities were measured at each point. By taking advantage of the $\Ct$ rotation symmetry of the moir\'e superlattices, the data were interpolated by a low order 2D Fourier expansion throughout configuration space, greatly reducing the number of calculations required. The stacking energy and equilibrium layer separation both vary by about 1 {\AA} / 0.1 eV, respectively, which is expected. We also found that both polarizability parameters vary significantly throughout the configuration space. Additionally, 3R \chem{MoS_2} has a small nonvanishing local dipole density with zero mean, see Fig.~\ref{fig:Fig2} (e), whereas 2H \chem{MoS_2} does not.

The lattice relaxation at finite electric fields was then performed in configuration space, including an additional in-plane displacement $\mathbf{u}(\mathbf{s})$ in the stacking and electrostatic energies, and minimizing the total energy numerically. The two terms in $\V_{\text{elec}}$ were included separately in order to illustrate their individual effects. The quadratic term does not break any symmetries, and thus we can study its effect on the domain structures by mapping the 2D moir\'e superlattice to a 1D Frenkel-Kontorova (FK) model\cite{popov2011commensurate,lebedeva2016dislocations,lebedeva2019commensurate}, with a single domain wall across the path AB $\to$ SP $\to$ BA (see Section III of the supplementary information for more details). The linear term breaks the $\mathcal{C}_6$ rotation symmetry of the 3R moir\'e superlattice, leading to a splitting between the AB and BA domains, which cannot be captured by a 1D FK model.

\begin{figure}[h]
\centering
\hspace*{-0.25cm}
\includegraphics[width=\columnwidth]{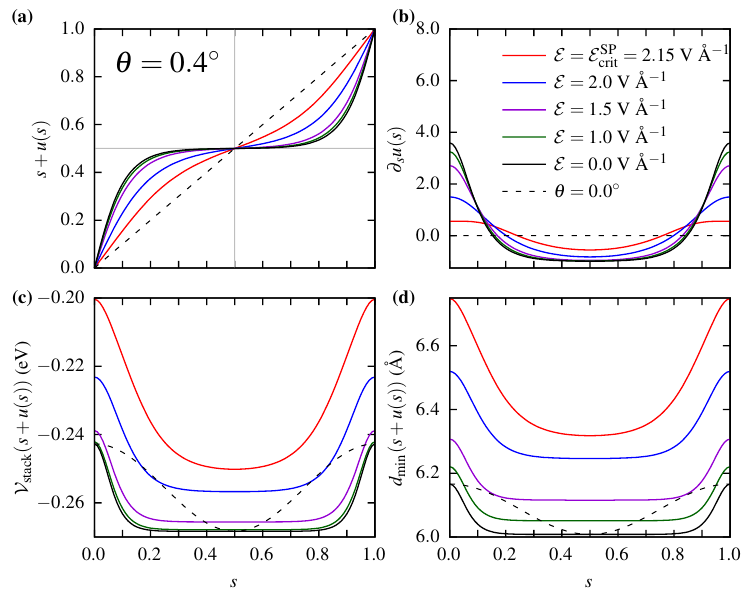}
\caption{Lattice relaxation from the 1D FK model, including the quadratic electrostatic term. \textbf{(a)}: total displacement in configuration space $s+u(s)$, \textbf{(b)}: gradient of displacement $\partial_su(s)$, \textbf{(c)}: stacking energy as a function of total displacement $V_{\text{stack}}(s+u(s))$ and \textbf{(d)}: equilibrium layer separation as a function of total displacement $\dmin(s+u(s))$.}
\label{fig:Fig3}
\end{figure}

\begin{figure*}[t]
\includegraphics[width=\textwidth]{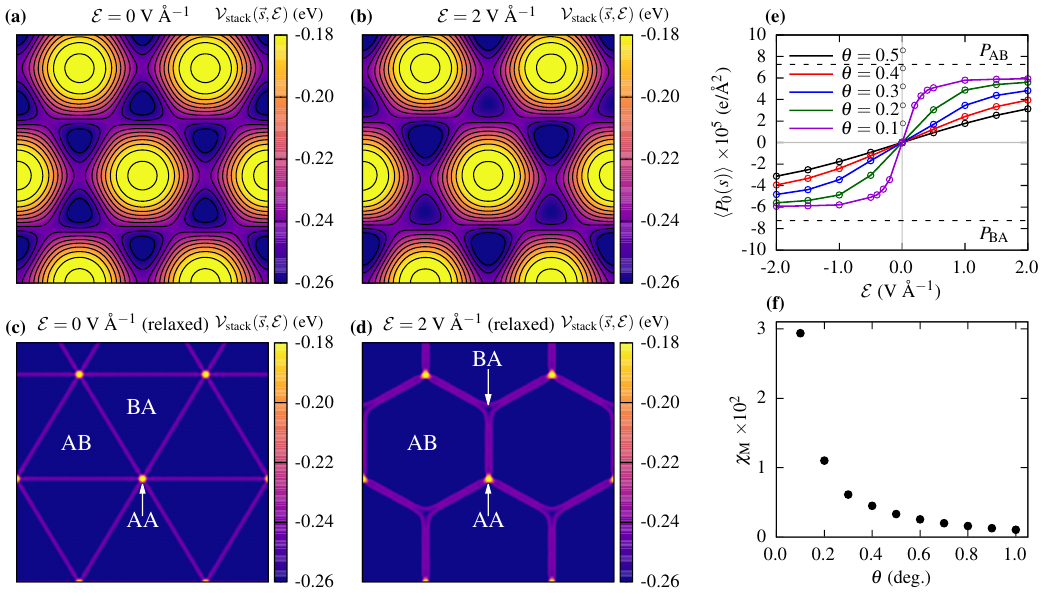}
\caption{Electrically tunable lattice relaxation and resulting polarization for 3R-stacked \chem{MoS_2}. \textbf{(a)-(d)}: Lattice relaxation in configuration space for 3R \chem{MoS_2} including the linear electrostatic term. The top panels show the stacking energy before lattice relaxation at electric field strengths of \textbf{(a)}: 0 V \AA$^{-1}$ and \textbf{(b)}: 2 V \AA$^{-1}$. \textbf{(c)} and \textbf{(d)} show the corresponding stacking energies after lattice relaxation at a twist angle of $\th = 0.1^{\circ}$. The AA regions (yellow) shrink and the AB/BA regions (purple) expand. When an electric field is applied, the AB and BA regions relax unevenly, one increasing in area and the other decreasing. \textbf{(e)}: average polarization as a function of electric field for several twist angles. \textbf{(f)}: Susceptibility of the moir\'e superlattice $\chi_{\text{M}}$ as a function of twist angle, obtained from the slopes from \textbf{(e)} about $\Ep=0$.}
\label{fig:Fig4}
\end{figure*}

The 1D FK model, including only the dielectric response to the field, was solved for a fixed lattice mismatch $\th$, and for several values of electric field, see Fig.~\ref{fig:Fig3}, with similar results at zero field for various lattice mismatches shown in Fig~S10. At zero field, decreasing $\th$ increases the displacement $u(s)$ in between AB/BA and SP points, leading to a domain structure with wide AB/BA regions and narrow SP regions, separated by a domain wall, with width proportional to $\th$ in configuration space. The domain structure is also evident from the stacking energy and equilibrium layer separation profile in configuration space, shown in Figs.~\ref{fig:Fig3} (c) and (d). When an electric field is applied, the equilibrium layer separation increases everywhere in configuration space, which decreases the stacking energy. With the stacking energy reduced, it is not as favorable for the atoms to relax; the displacement decreases, and the domain walls soften. 

The 2D relaxation including only the coupling between the field and spontaneous polarization was done for a range of twist angles $0.1^{\circ} \leq \th \leq 1.0^{\circ}$ and field strengths $0 \leq \Ep \leq 2$ V \AA$^{-1}$. The results are summarized in Fig.~\ref{fig:Fig4}, and additional plots for all angles and electric field strengths can be found in Section IV of the supplementary information. In Figs~\ref{fig:Fig4} (a) and (b) we show the stacking energy in 3R \chem{MoS_2} in configuration space obtained from first-principles calculations, at electric field strengths of 0 and 2 V \AA$^{-1}$, respectively. The electric field increases the depth of the well at AB and decreases the depth of the well in BA by the same amount, breaking the $\Cs$ rotation symmetry. The panels below, Figs~\ref{fig:Fig4} (c) and (d), show the corresponding stacking energies after lattice relaxation for a twist angle of $\th = 0.1^{\circ}$. At zero field, the relaxation reduces the area of the AA regions and increases the area of the AB/BA regions, leading to a triangular domain structure with sharp domain walls. When an electric field is applied, the AB and BA regions relax unevenly, leading to larger AB regions and smaller BA regions reducing the rotation symmetry to $\Ct$. When the AB and BA domains are no longer equal in area, the polarization no longer averages to zero. In Fig.~\ref{fig:Fig4} (e) we show the average spontaneous polarization $\left<P_0(s)\right>$ in configuration space as a function of field for different twist angles. We can see that the response to the field is very sensitive to the twist angle. In Fig.~\ref{fig:Fig4} (f) we show the susceptibility of the moir\'e superlattice $\chi_{\text{M}}$ as a function of twist angle, which was obtained by taking the slope of the polarization about zero field. We can see that the susceptibility increases dramatically as the twist angle decreases.

\section*{Discussion}

We have introduced a model which illustrates the effect of an applied electric field on lattice relaxation in moir\'e superlattices. The model contains two electrostatic contributions. The first is a linear coupling between the field and the local spontaneous polarization in bilayers without centrosymmetry, which breaks the degeneracy between the AB and BA stacking domains. Under an electric field, the AB and BA regions will relax unevenly with one growing and the other shrinking with respect to the relaxed structure at zero field. This leads to a nonzero average out-of-plane polarization in the superlattice. The second contribution is the dielectric response to the field, which occurs in all bilayers. This term leads to a nonuniform increase the layer separation, which reduces stacking energy, leading to softer domains structures under lattice relaxation.

Finally, as our theory does not predict a ferroelectric response, we offer some thoughts on the recent experimental observations of ferroelectricity in the context of our  model. 

For a system to be considered ferroelectric, it  (i) must  exhibit a spontaneous polarization at zero field which (ii) must be switchable with an electric field. However, neither the individual stacking domains nor the moir\'e superlattice as a whole satisfy both conditions: The stacking domains indeed have a local spontaneous polarization at zero field, and while the average polarization of a domain can change via lattice relaxation under an electric field, the sign of the polarization in each cannot be switched. Therefore, the stacking domains in moir\'e superlattices are in general not ferroelectric. Conversely, the moir\'e superlattice itself exhibits an average polarization, the direction of which can be changed by the field, but has zero average polarization at zero field. Therefore, under ideal conditions, moir\'e superlattices are also not ferroelectric. 

This idealized picture may not hold in experimental settings, and defects, mislocations or strain induced by the finite size of samples may lead to uneven domains at zero field. Also, the direct couplings between strain and polarization, piezoelectricity and flexoelectricity, have not been considered, which may make it energetically favourable for the domains to relax unevenly and the superlattice to have a nonzero average polarization at zero field. We leave the consideration of these effects for future work. To summarize, for an ideal system, when considering the spontaneous local polarization and lattice relaxation under an electric field, \textit{neither the moir\'e superlattice nor the stacking domains are ferroelectric}, since the former does not have a spontaneous polarization at zero field and the latter does not have a switchable polarization.

There have also been reports of a switching of the polarization in a single stacking domain by a sliding of the atoms by half a monolayer lattice constant when a local field was applied to the domain using a biased atomic force microscopy (AFM) tip \cite{stern2020interfacial}. This sliding change the stacking configuration: AB $\leftrightarrow$ BA, leading to a first order switching of the polarization. This is a separate mechanism to the one mediated by lattice relaxation, which results in a second order change in the polarization. We can understand this sliding in the context of our model. When a field is applied to the domain, the linear coupling between the field and polarization will lower the energy if the two are aligned and result in a large energy penalty if they are anti-aligned. In either case, the dielectric response will reduce the stacking energy by the same amount, making it easier for one layer to slide with respect to the other. When the field and polarization are anti-aligned, the energy can be lowered considerably via a sliding by half a monolayer unit cell, flipping the polarization so that it becomes aligned with the field. However, the field is applied to the domain locally via an AFM tip, and it is not clear whether the sliding occurs locally under the tip, or throughout the entire domain. It is also not clear whether or not the domain will remain flipped once the field is removed, or relax back to its original orientation. Thus, it is not clear whether or not this mechanism for a first order switching of polarization in a stacking domain is truly ferroelectric either.

The model introduced in this paper illustrates, clearly and intuitively, the effect an electric field can have on lattice relaxation in moir\'e superlattices. We propose an electric field as a third quantity with which the domain structures in moir\'e superlattices can be tuned. Unlike the twist angle and lattice mismatch which are fixed for a given sample, an electric field can be applied dynamically to tune a sample in-situ. Thus, it may serve as a more practical approach to achieve control in moir\'e superlattices. We have also discussed how our theoretical model can be used to understand the recent observations of ferroelectricity in moir\'e superlattices. We believe it is inaccurate to consider moir\'e materials to be truly ferroelectric via lattice relaxation or sliding under an electric field. However, this motivates further study into polar phenomena in moir\'e materials.

\section*{Methods}

\subsection*{First-principles calculations}

First-principles density functional theory (DFT) calculations were performed using the {\sc siesta} code\cite{siesta} using PSML\cite{psml} norm-conserving\cite{norm_conserving} pseudopotentials, obtained from pseudo-dojo\cite{pseudodojo}. {\sc siesta} employs a basis set of numerical atomic orbitals (NAOs)\cite{siesta,siesta_2}, and double-$\zeta$ polarized (DZP) orbitals were used for all calculations. The basis sets were optimised by hand, following the methodology in Ref. [\onlinecite{basis_water}].

A mesh cutoff of $1200 \ \text{Ry}$ was used for the real space grid in all calculations. A Monkhorst-Pack $k$-point grid\cite{mp} of $12 \times 12 \times 1$ was used for the initial geometry relaxations, and a mesh of $18 \times 18 \times 1$ was used to calculate polarizabilities. Calculations were converged until the relative changes in the Hamiltonian and density matrix were both less than $10^{-6}$. For the geometry relaxations, the atomic positions were fixed in the in-plane directions, and the vertical positions and in-plane stresses were allowed to relax until the force on each atom was less than $0.1 \ \text{meV \AA$^{-1}$}$. The layer separation $d$ was taken to be the distance between the carbon atoms in bilayer graphene an the distance between the metals in bilayer \chem{MoS_2} (see Fig~\ref{fig:Fig1} (e)), and the stacking energy is calculated as $\V_{\text{stack}} = \V_{\text{bilayer}} - 2\V_{\text{mono}}$, where $\V_{\text{bilayer}}$ and $\V_{\text{mono}}$ are the total energies of the bilayer and monolayer systems, respectively. The polarizability to zeroth order in $d$, $\a_0$, was obtained by fixing the relaxed geometry, applying an electric field large enough to overcome internal field effects, and measuring the change in the out-of-plane dipole moment of the bilayer. The polarizability to first order in $d$, $\a_1$, was obtained by changing the layer separation by $\pm 1 \%$ with respect to $d_0$ and measuring the relative change in the polarizability. A detailed first-principles study is provided in Section II of the supplementary information.

\subsection*{Lattice Relaxation}

Eq.~\eqref{eq:V_tot_main} can be minimized by using variational methods and solving the resulting differential equations. This can be relatively demanding in 2D. Instead, we minimized the total energy using numerical optimization methods. Eq.~\eqref{eq:V_tot_main} is a functional of the in-plane displacement $\mathbf{u}$ in configuration space. If we perform a plane-wave expansion, $\mathbf{u}(\mathbf{s}) = \sum_{\mathbf{G}} \mathbf{u}_{\mathbf{G}} e^{i \mathbf{G}\cdot\mathbf{s}}$, where $\mathbf{G}$ are the reciprocal lattice vectors of the commenusrate bilayer, then the total energy becomes a function of $\{\mathbf{u}_{\mathbf{G}}\}$, and can be minimized numerically with respect to the coefficients: $\grad_{\mathbf{U}_{\mathbf{G}}}V_{\text{tot}} = 0$. This was done in {\sc Julia}, using the {\sc Optim} package to do the optimization.

We can take advantage of the $\Ct$ symmetry of our model to greatly reduce the number of independent $\mathbf{G}$ vectors:
\beq{}
\begin{split}
\mathbf{u}(\mathbf{s}) = \sum_{n = 1}^3\sum_{\mathbf{G}} \Ct^{n-1}& \left[(\mathbf{u}_{\mathbf{G}}+\mathbf{u}_{-\mathbf{G}})\cos{\lb \Ct^{n-1} \mathbf{G}\cdot\mathbf{s}\rb } \right.\\
&\left. + i (\mathbf{u}_{\mathbf{G}}-\mathbf{u}_{-\mathbf{G}})\sin{\lb \Ct^{n-1} \mathbf{G}\cdot\mathbf{s}\rb }\right].
\end{split}
\eeq
When there is a $\Cs$ symmetry, i.e.~for 3R \chem{MoS_2} at zero field, we have $\mathbf{u}_{-\mathbf{G}} = - \mathbf{u}_{\mathbf{G}}$, and the cosine terms vanish. The optimization was done using the independent $\mathbf{G}$ in the first five shells (10 vectors) for $\th \geq 0.5^{\circ}$, six shells (21 vectors) for $0.5^{\circ} > \th \geq 0.3^{\circ}$ and and seven shells (28 vectors) for $0.3^{\circ} >\th \geq 0.1^{\circ}$ . The total energy was optimized using the limited memory BFGS algorithm (L-BFGS), until the gradient was below $1\times 10^{-5} \ \text{eV}$.\\

\noindent \textbf{Data Availability.} The data presented were generated from first-principles and lattice relaxation calculations as described in the text.

\noindent \textbf{Code Availability.} Code is available upon reasonable request.

\noindent \textbf{Acknowledgements.} D.~B would like to thank E.~Artacho and I.~Lebedeva for helpful discussions. D.~B acknowledges funding from the EPSRC Centre for Doctoral Training in Computational Methods for Materials Science under grant number EP/L015552/1. B.~R gratefully acknowledges support from the Cambridge International Trust.

\noindent \textbf{Author Contributions.} D.~B. conceived the project and performed the first-principles calculations. D.~B. and B.~R. developed the theoretical model and performed the lattice relaxation calculations. D.~B. wrote the manuscript, with contributions from B.~R.

\noindent \textbf{Competing Interests.} The authors declare no competing interests.

\end{document}

%%% TITLE, AUTHORS, ABSTRACT%%%

\title{SUPPLEMENTARY INFORMATION \\ On electrically tunable stacking domains and ferroelectricity in moir\'e superlattices}

\author{Daniel Bennett}
%\email{db729@cam.ac.uk}
 \affiliation{Theory of Condensed Matter, Cavendish Laboratory, Department of Physics, J J Thomson Avenue, Cambridge CB3 0HE, United Kingdom}
 
\author{Benjamin Remez}
\affiliation{Theory of Condensed Matter, Cavendish Laboratory, Department of Physics, J J Thomson Avenue, Cambridge CB3 0HE, United Kingdom}

\date{\today}

\maketitle

\section*{Supplementary Note 1: Physical Model}

\subsection*{Twistronics}
\begin{figure}[h]
\includegraphics[width=0.75\linewidth]{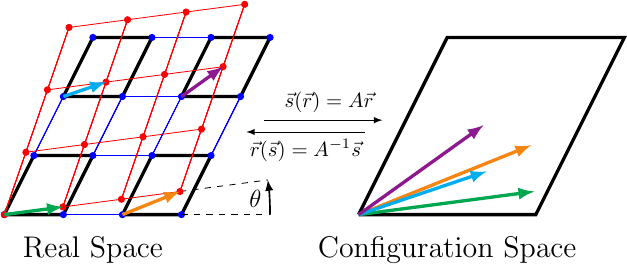}
\caption{Sketch of the mapping between real space and configuration space. On the left is a $3\times 3$ section of a moir\'e superlattice with twist angle $\th$. The positions $\mathbf{r}$ are mapped into configuration space via the operator $A = (I - R_{\th}^{-1})$, modulo a primitive unit cell. The vectors $\mathbf{s}$ are shown for the four highlighted cells.}
\label{fig:diagram-slide}
\end{figure}
A twisted bilayer system is composed of two layers with a relative twist angle $\th$ between them, which we call the reference (r) and twisted (t) layers, respectively. The lattice vectors of the reference layer and the twisted layer are
\beq{}
\begin{split}
\mathbf{a}_{\text{r},1} &= a\begin{bmatrix} 1 \\ 0 \end{bmatrix}, \qquad \mathbf{a}_{\text{r},2} = \frac{a}{2}\begin{bmatrix} 1 \\ \sqrt{3} \end{bmatrix}\\
\mathbf{a}_{\text{t},1} &= R_{\th} \mathbf{a}_{\text{r},1},\qquad \mathbf{a}_{\text{t},2} = R_{\th}\mathbf{a}_{\text{r},2}\\
\end{split}
\eec
respectively, where $a$ is the monolayer lattice constant and $R_{\theta} = \begin{bmatrix} \cos{(\th)} & -\sin{(\th)} \\ \sin{(\th)} & \cos{(\th)} \end{bmatrix}$. For a general $\th$, the two layers are incommensurate, i.e.~they form a supercell which is infinitely large. If the two layers form a commensurate supercell we can define supercell lattice vectors $\mathbf{L}_i$ as a linear combinations of the lattice vectors of either layer:
\beq{eq:t_supercell}
\begin{split}
\mathbf{L}_1 &= p \mathbf{a}_{\text{r},1} + q \mathbf{a}_{\text{r},2}\\
    &= \bar{p} \mathbf{a}_{\text{t},1} + \bar{q} \mathbf{a}_{\text{t},2} = \bar{p} R_{\th} \mathbf{a}_{\text{r},1} + \bar{q} R_{\th} \mathbf{a}_{\text{r},2}\\
\mathbf{L}_2 &=  R_{\pi/3} \mathbf{L}_1
\end{split}
\eep
The second line of Eq.~\eqref{eq:t_supercell} leads to a Diophantine equation in the integers $p$, $q$, $\bar{p}$ and $\bar{q}$, and the set of angles which result in a commensurate supercell is given by\cite{shallcross2010electronic,mele2010commensuration,hermann2012periodic}
\beq{}
\th(p,q) = \cos^{-1}{\lb \frac{p^2 + 4pq + q^2}{2 \lb p^2 + pq + q^2\rb}\rb}
\eep
Consider an atom in the reference layer, with in-plane position $\mathbf{r}_0$. The corresponding atom in the twisted layer has position $\mathbf{r} = R_{\th}\mathbf{r}_0 $. The displacement due to twisting is ${\mathbf{\d}(\mathbf{r}) = \mathbf{r}-\mathbf{r}_0 = \lb I - R_{\th}^{-1}\rb\mathbf{r}}$. If the layers form a commensurate supercell with lattice vectors $\mathbf{L}^{\text{M}}_i$, say, the atoms will realign when the displacement is equal to a lattice vector, $\d(\mathbf{L}^{\text{M}}_i) = \mathbf{a}_{\text{r},i}$:
\beq{}
\mathbf{L}^{\text{M}}_i = \lb I - R_{\th}^{-1}\rb^{-1}\cdot \mathbf{a}^1_i
\eep
The cell spanned by $\mathbf{L}^{\text{M}}_i$ is known as a moir\'e superlattice, and
\beq{}
L_{\text{M}} \equiv \left| \mathbf{L}^{\text{M}}_i\right| = \frac{a}{2\sin{(\th/2)}}
\eec
is known as the the moir\'e period, which is is not necessarily equal to the supercell period:
\beq{}
L_{\text{sc}} = \frac{\left|p-q\right|a}{2\sin{(\th/2)}} = \left|p-q\right|L_{\text{M}}
\eep
For $p\neq q$, there will be $(p-q)^2$ moir\'e periods in the supercell.

Having established a geometric description, we can model structural or electronic phenomena in a twisted bilayer using phenomenological or tight-binding models. These models can be paramaterized using first-principles calculations. However in practice this is difficult to do in real space because the size of and the number of atoms in the supercell becomes prohibitively large at smaller twist angles. Fortunately, we can take advantage of a useful mapping which allows us to parameterize systems at arbitrary twist angles using just a single cell of a commensurate bilayer.

The set of displacements of every atom in a moir\'e superlattice can be described by $\mathbf{\d}(\mathbf{r})$. Alternatively, we can describe the displacements by a set of local translations
\beq{eq:slide_map}
\begin{split}
\mathbf{s}(\mathbf{r}) &= \lb I - R^{-1} \rb \mathbf{r} \mod \{\mathbf{a}_{\text{r},1},\mathbf{a}_{\text{r},2}\} \\ 
& \equiv A\cdot \mathbf{r}
\end{split}
\eec
i.e.~at $\mathbf{r}$, the system is locally equivalent to an untwisted bilayer, with a relative in-plane slide $\mathbf{s}_{\th}(\mathbf{r})$ between the layers, see Fig.~\ref{fig:diagram-slide}. The set of translations $\mathbf{s}(\mathbf{r})$ is contained in a single primitive cell of the reference layer, even in the continuum limit. We call this space of translations configuration space\cite{carr2017twistronics,cazeaux2017analysis,massatt2017electronic,carr2018relaxation}. This space can be traversed by taking an untwisted commensurate bilayer and sliding one layer over the other. Physical properties can be measured in configuration space, and we can use the inverse map,
\beq{eq:slide_map_inverse}
\mathbf{r}(\mathbf{s}) = \lb I - R_{\th}^{-1}\rb^{-1} \mathbf{s} = A^{-1} \mathbf{s}
\eec
to parametrize a moir\'e superlattice in real space for arbitrary twist angles\cite{carr2018relaxation}. Derivatives in real space and configuration space are related by
\beq{eq:slide_map_derivative}
\grad_r = A^T \grad_s
\eec
which is useful for mapping quantities which depend on spatial derivatives, such as strains, to configuration space.

\subsection*{Stacking energy}

It is well known layered systems interact via long-range van der Waals forces. For non-polar materials, this is facilitated by induced dipole$-$induced dipole, or London interactions, where fluctuations in the charge density of one layer lead to a dipolar response in the other layer, and vice versa. This can be described by a van der Waals potential in the layer separation $d$:
\beq{eq:V_stacking_1}
\V_{\text{stack}}(d) = \frac{n}{m-n}\left|\V_0\right|\left[\lb\frac{d_0}{d}\rb^{m}-\frac{m}{n}\lb\frac{d_0}{d}\rb^n \right]
\eec
where $\V_0$ is the depth of the potential well or cohesive energy \textit{per unit cell}, $d_0$ is the equilibrium separation, and the indices $(n,m)$ determine the curvature of the well about the minimum. Eq.~\eqref{eq:V_stacking_1} can be parametrized using first-principles calculations, which is done in Appendix B. For two slabs which extend infinitely in area, the energy is expected to behave like $d^{-2}$ at larger separations, since the dipole-dipole interaction is integrated twice over an infinite area\cite{tadmor2001london}. From first-principles calculations we found that this is not the case, and the long-range interactions in bilayer graphene and \chem{MoS_2} decay with $n\geq 6$. We suspect that this is because a bilayer is not adequately described as a pair of capacitor plates, but rather a single slab with a non-uniform charge density as in Fig.~1 (e). The non-zero overlap of states in the vacuum region may be screening the long-range interactions and could be responsible for larger than expected values of $(n,m)$, although a more detailed study is required to verify this.

\subsection*{Electrostatic energy}

The electrostatic energy of a dielectric slab in the presence of a perpendicular electric field $\Ep$ is
\beq{eq:V_elec_1}
\V_{\text{elec}}(P,\Ep,d) = \Omega \lb \frac{1}{2\eo\x(d)}(P-P_0)^2 - \Ep P\rb
\eec
where $P$ and $P_0$ are the total and spontaneous polarization, respectively, $A$ is the in-plane area of the bilayer, $\Omega = Ad$ is the volume of the bilayer, and $\x(d)$ is the dielectric susceptibility, which in general depends on the geometry of the system. Eq.~\eqref{eq:V_elec_1} assumes a linear response to the applied field, $P(\Ep,d) = P_0 + \eo\x(d)\Ep$, which inserting into Eq.~\eqref{eq:V_elec_1} gives:
\beq{eq:V_elec_2}
\V_{\text{elec}}(\Ep,d) = -\frac{1}{2}\eo\a(d)\Ep^2 - \Ep p_0
\eec
where $\a = \Omega\x$ is the polarizability, which describes the linear response of the dipole moment to the applied field, and $p_0 = \Omega P_0$ is the spontaneous dipole moment. When an electric field is applied, the layers are no longer non-polar, and the long-range interactions can no longer be considered induced dipole$-$induced dipole interactions. However, we can think of the interactions between the layers as arising due to fluctuations about a finite polarization, rather than zero polarization.

The effect of an applied field can be estimated from first-principles by measuring the stacking energy as a function layer separation for different field strengths. In Ref.~\onlinecite{santos2013electrically}, it was found that the electric field lowers the stacking energy of bilayer \chem{MoS_2} for large separations, leading to a breakdown of the bilayer at $\Ep \sim 2$ \text{V \AA$^{-1}$}, although the precise behavior of the equilibrium layer separation as a function of applied field is unclear. In order to clarify this, we performed detailed first-principles calculations in Appendix B. In Fig.~\ref{fig:DFT-V-Q} (a) we plot the potential energy curves for bilayer \chem{MoS_2} as a function of $d$ for different values of applied field. We see that the electric field lowers the energy at larger separations, and the bilayer becomes unstable at $\Ecrit \approx 2.25 \ \text{V \AA}^{-1}$, similar to the results obtained in Ref.~\onlinecite{santos2013electrically}. By calculating the M{\"u}lliken charges on each layer in Fig.~\ref{fig:DFT-V-Q} (b), we find that there is an interlayer charge transfer when an electric field is applied. The charge transfer is linear in $\Ep$ above $\Ep \approx 0.27 \ \text{V \AA$^{-1}$}$, below which little or no charge transfer is observed. We attribute this to internal field effects, similar to the depolarizing field observed in ferroelectric thin films. see Fig.~1 (d). The internal field could be accounted for in the electrostatic energy as\cite{santos2013electrically} $\V_{\text{int}} = -\frac{1}{2}\Ep_{\text{int}} P$. In the interest of simplicity, we neglect internal field effects in our model.

\begin{figure}[t]
\centering
%\hspace*{-0.5cm}
\includegraphics[width=0.65\linewidth]{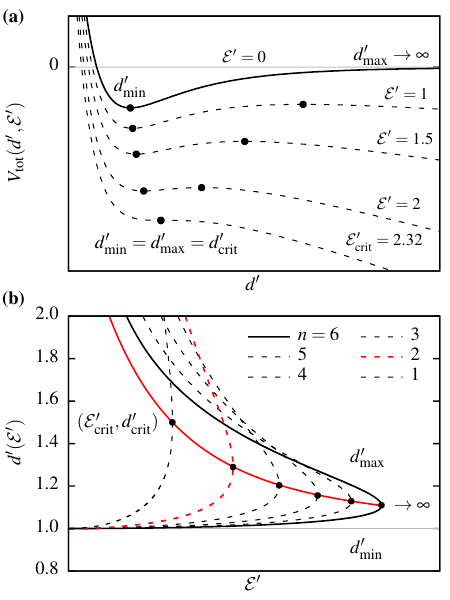}
\caption{Analytic results from Eq.~\eqref{eq:V_stacking_elec_1}. \textbf{(a)}: Total energy from Eq.~\eqref{eq:V_stacking_elec_1} as a function of $d$ for increasing values of $\Ep$. The points show the positions of $\dmin$ and $\dmax$ on each curve. The lowest curve is for field strength of $\Ep = \Ecrit$, where $\dmin = \dmax \equiv \dcrit$. \textbf{(b)}: $\dmin$ and $\dmax$ as a function of $\Ep$ for different values of $n$, increasing from left to right. The solid black curve is $n=6$, and the dashed red curve is the theoretical value $n=2$. The solid red curve shows the critical values $(\dcrit,\Ecrit)$ from Eq.~\eqref{eq:d_E_crit}. The critical points on each curve are marked, i.e.~when $d=\dcrit$ and the bilayer breaks down.}
\label{fig:model-V-d}
\end{figure}

In Fig.~\ref{fig:DFT-p-alpha} (a) we plot the dipole moment $p(\Ep)$ for several values of $d$. For field strengths strong enough to overcome internal field effects, the dipole moment is linear in $\Ep$, but also in $d$. Thus, as shown in  Fig.~\ref{fig:DFT-p-alpha} (b), the polarizability is approximately linear in $d$,
\beq{eq:alpha}
\a(d) = \a_0 + \a_1 \lb \frac{d}{d_0}-1\rb
\eec
where $\a_0$ and $\a_1$ are both positive and have units of \AA$^3$. Taking $(n,m) = (n,2n)$, we can write the energy, exlcuding elastic contributions, as
\beq{eq:V_stacking_elec_1}
\begin{split}
\V_{\text{tot}}(d,\Ep) =& \left|\V_0\right|\left[\lb\frac{d_0}{d}\rb^{2n}-2\lb\frac{d_0}{d}\rb^n \right] \\
&- \Ep p_0 - \frac{1}{2}\eo\lb \a_0 + \a_1\lb \frac{d}{d_0}-1\rb\rb\Ep^2
\end{split}
\eep
By plotting contours of the energy as a function of $d$ for fixed values of $\Ep$, see Fig.~\ref{fig:model-V-d} (a), we can see that the stacking energy is lowered by the electric field, consistent with results from first-principles calculations. At zero field, $\V_{\text{stack}}$ has a minimum at the equilibrium separation $d_{\text{min}} = d_0$, and a maximum (for $d>d_{\text{min}}$) at $d_{\text{max}}\to\infty$. When a field is applied, $\V_{\text{stack}}$ diverges as $d\to\infty$. This makes physical sense: increasing $d$ will increase the total dipole moment and lower the total energy. For a non-zero field, $d_{\text{max}}$ has a finite value, at the top of the energy barrier which separates $d_{\text{min}}$ and $d\to\infty$. As the field strength increases, $d_{\text{min}}$ and $d_{\text{max}}$ move closer together, eventually meeting at a critical point $(\dcrit, \Ecrit)$ where the energy barrier vanishes, and the bilayer becomes unstable,
\beq{eq:d_E_crit}
\begin{split}
d_{\text{crit}}' &= \lb\frac{2n+1}{n+1}\rb^{\frac{1}{n}}\\
\Ecrit' &= 2n\sqrt{\frac{(n+1)^{1+\frac{1}{n}}}{(2n+1)^{2+\frac{1}{n}}}}
\end{split}
\eec
where $d'\equiv\frac{d}{d_0}$ and $\Ep'\equiv\sqrt{\frac{\eo\a_1 d_0}{\left|\V_0\right|}}\Ep$. In Fig.~\ref{fig:model-V-d} (b) we show $d_{\text{min}}$ and $d_{\text{max}}$ as a function of $\Ep$ for several values of $n$. The solid red line shows Eq.~\eqref{eq:d_E_crit}, which separates $\dmin$ and $\dmax$. Interestingly, the behavior is highly sensitive to the value of $n$. For larger values of $n$, which are predicted from first-principles calculations, $d_{\text{min}}$ does not change by much until $\Ep \sim \Ecrit$, where it increases by about 10\% before the bilayer becomes unstable. For smaller values of $n$, the layers separate more easily because the potential well is shallower, but the breakdown occurs at smaller field strengths.

\subsection*{Elastic energy}

The elastic energy in Eq.~2 is given as a contraction of the strain tensors with the fourth rank elasticity tensor \cite{carr2018relaxation,zhu2020twisted}:
\beq{}
\begin{split}
C_{11} = \begin{bmatrix} B+\mu & 0 \\ 0 & B-\mu\end{bmatrix}, \quad C_{12} = \begin{bmatrix} 0 & \mu \\ \mu & 0\end{bmatrix},\\
C_{21} = \begin{bmatrix} 0 & \mu \\ \mu & 0\end{bmatrix}, \quad C_{22} = \begin{bmatrix} B-\mu & 0 \\ 0 & B+\mu\end{bmatrix}
\end{split}
\eeq
and is given explicitly as 
\beq{}
\begin{split}
\V_{\text{elastic}} &= \frac{1}{2} \left[ B\lb \ep_{11} + \ep_{22}\rb^2 + \mu \lb \lb \ep_{11}-\ep_{22} \rb^2 + \lb \ep_{12} + \ep_{21}\rb^2 \rb \right]\\
&= \frac{1}{2} \left[ B\lb \partial_xU_x + \partial_yU_y\rb^2 + \mu \lb \lb \partial_xU_x - \partial_yU_y \rb^2 + \lb  \partial_xU_y + \partial_yU_x\rb^2 \rb \right]
\end{split}
\eeq
The strain tensor in configuration space is\cite{carr2018relaxation}
\beq{}
\ep_{ij} = \frac{1}{2}\lb (\partial_i u_k)A_{kj} + (\partial_j u_k)A_{ki}\rb
\eeq
which allows us to write down the elastic energy in configuration space:
\beq{}
\begin{split}
\V_{\text{elastic}} &= \frac{\th^2}{2} \left[ B\lb \partial_{s_x}u_{s_y} - \partial_{s_y}u_{s_x}\rb^2 + \mu \lb \lb \partial_{s_x}u_{s_y} + \partial_{s_y}u_{s_x} \rb^2 + \lb  \partial_{s_x}u_{s_x} - \partial_{s_x}u_{s_x}\rb^2 \rb \right]
\end{split}
\eeq
It is convenient to work in terms of the lattice vectors $\mathbf{a}_{\text{r},1}$ and $\mathbf{a}_{\text{r},2}$. Under this transformation, the displacement transforms as $\mathbf{u}\to g\mathbf{u}$ and the strain tensor transforms as ${\ep_{ij} \to g^{-1}_{i\a}\ep_{\a\b} g_{\b j} }$, where $g = \begin{bmatrix} 1 & 1/2 \\ 0 & \sqrt{3}/2\end{bmatrix}$. The elastic energy is then 

\beq{}
\begin{split}
\V_{\text{elastic}} &= \frac{\th^2}{2} \left[ B\lb \partial_{a_x}u_{a_y} - \partial_{a_y}u_{a_x}\rb^2 + \mu \lb \frac{4}{3}\lb \partial_{a_x}u_{a_y} + \partial_{a_y}u_{a_x} \rb^2 + \lb  \partial_{a_x}u_{a_x} - \partial_{a_y}u_{a_y}\rb^2 \rb \right]
\end{split}
\eeq

\clearpage

\section*{Supplementary Note 2: First-principles calculations}

\begin{figure}[h]
\centering
%\hspace*{-0.25cm}
\includegraphics[width=0.65\linewidth]{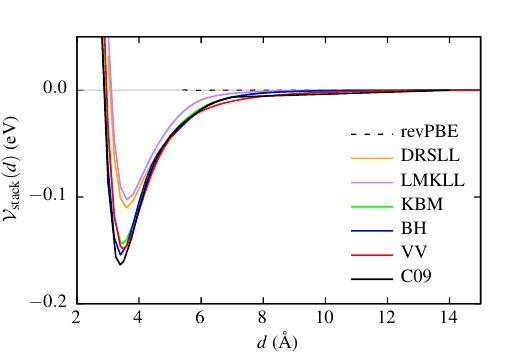}
\caption{Stacking energy as a function of layer separation for bilayer graphene for various VDW-corrected functionals in {\sc siesta}\cite{siesta}: revPBE\cite{zhang1998comment} (GGA), DRSLL (DF1)\cite{dion2005erratum,roman2009efficient}, LMKLL (DF2) \cite{dion2005erratum,lee2010higher}, KMB \cite{dion2005erratum,klimevs2009chemical}, C09 \cite{dion2005erratum,cooper2010van}, BH \cite{dion2005erratum,berland2014exchange}, VV \cite{vydrov2010nonlocal}.}
\label{fig:DFT-vdw-functional}
\end{figure}

Because functionals within the generalized gradient approximation (GGA) tend to underestimate the cohesive energy of van der Waals systems, the different VDW-corrected functionals in {\sc siesta} were tested by measuring the stacking energy of bilayer graphene as a function of layer separation, see Fig.~\ref{fig:DFT-vdw-functional}. The rest of the calculations were performed using the C09 functional, as it was found to give good results for both bilayer graphene and \chem{MoS_2}.

\subsection*{Parametrizing the stacking energy}

\begin{figure}[h]
\centering
%\hspace*{-0.25cm}
\includegraphics[width=0.65\linewidth]{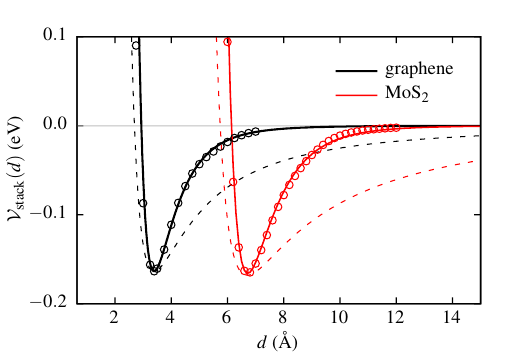}
\caption{Parameterization of $\V_{\text{stack}}(d)$ for bilayer graphene and \chem{MoS_2} (3R stacking). The points show results from DFT calculations, and the solid lines show the parameterization using the values in Table \ref{table:dft}. The dashed lines show the same parameterization, but with the smaller index changed to 2.}
\label{fig:DFT-vdw-graphene}
\end{figure}

We can parameterize the stacking energy by calculating $\V_0$, $d_0$ and the two indices $(n,m)$. Using suitable values for $(n,m)$, a fit to the stacking energy as a function of layer separation was obtained using Eq.~\eqref{eq:V_stacking_1}, which is shown in Fig.~\ref{fig:DFT-vdw-graphene} for bilayer graphene and \chem{MoS_2} with 3R stacking. The parameters used to fit Eq.~\eqref{eq:V_stacking_1} to the first-principles calculations are given in Table~\ref{table:dft}.

\begin{table}[h]
\renewcommand*{\arraystretch}{1.5}
\setlength{\tabcolsep}{7pt}
\centering
%\hspace*{-0.15cm}
\begin{tabular}{c | c c c c }
\hline\hline
Material & $a \ (\text{\AA})$ &$\V_0 \ (\text{eV})$ & $d_0 \ (\text{\AA})$ & $(n,m)$ \\ \hline
graphene & 2.473 & -0.163 & 3.393 & $(6,8)$ \\ \hline
\chem{MoS_2} (3R) & 3.164 & -0.165 & 6.716 & $(8,15)$ \\ \hline \hline
\end{tabular}
\caption{Parameterization of $\V_{\text{stack}}(d)$ for bilayer graphene and \chem{MoS_2} from first-principles calculations.}
\label{table:dft}
\end{table}

\subsection*{Parametrizing the electrostatic energy}

To parametrize $\V_{\text{elec}}$, an electric field was applied in the out-of-plane direction. A dipole correction\cite{dipole_correction_1,dipole_correction_2,dipole_correction_3,dipole_correction_4} was used in the vacuum region to prevent long-range interactions between periodic images. The resulting dipole moment in the out-of-plane direction was then measured. 

\begin{figure}[h]
\centering
%\hspace*{-0.5cm}
\includegraphics[width=0.65\linewidth]{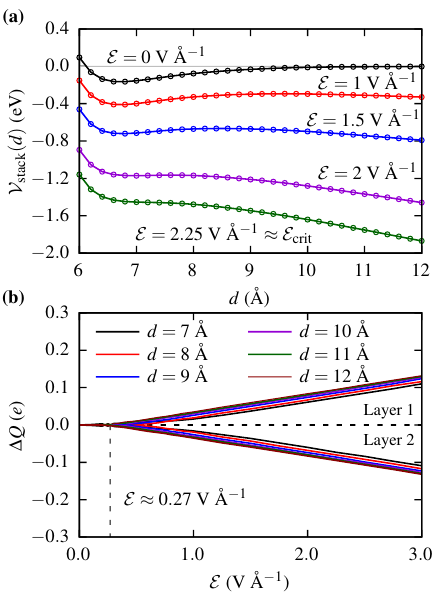}
\caption{Stacking energy and M\"ulliken charges on each layer as a function of $d$ and $\Ep$. \textbf{(a)}: $\V_{\text{stack}}(d)$ for different fixed values of electric field. \textbf{(b)}: Change in M{\"u}lliken charges $\D Q$ of each layer as a function of electric field. The vertical dashed line indicates the field strength beyond which internal field effects are mitigated.}
\label{fig:DFT-V-Q}
\end{figure}

We first performed geometry relaxations of bilayer \chem{MoS_2} for electric fields of increasing strength. We found that the layer separation increases only marginally (a very low force tolerance is required to see this, otherwise the layers do not move) until around $\Ep \sim 2 \ \text{V \AA$^{-1}$}$, where the layers begin to separate and the bilayer quickly becomes unstable. In order to investigate this peculiar behaviour, we calculated the stacking energy curves for various electric field strengths, see Fig.~\ref{fig:DFT-V-Q} (a). We can see that the electric field lowers the stacking energy, in agreement with our model. In Fig.~\ref{fig:DFT-V-Q} (b) we show the M{\"u}lliken chares as forces on each layer as a function of electric field, which indicates that there is a transfer of charge across the layers as an electric field is applied, verifying similar results obtained in Ref.~\onlinecite{santos2013electrically}.

\begin{figure}[h]
\centering
%\hspace*{-0.25cm}
\includegraphics[width=0.65\linewidth]{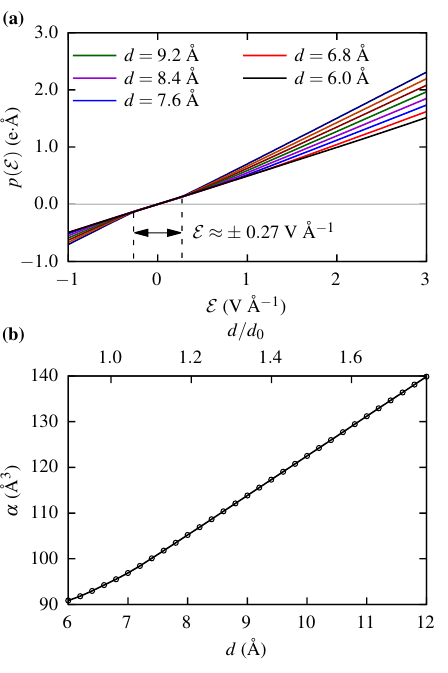}
\caption{Out-of-plane dipole moment and polarizability as a function of $d$ and $\Ep$. \textbf{(a)}: Out-of-plane dipole moment $p(\Ep)$ for various fixed values of $d$. The range of field values for which internal field effects are significant is indicated by the dashed lines. \textbf{(b)}: Polarizability $\a(d)$, measured at $\Ep > 0.27 \ \text{V \AA$^{-1}$}$ in order to neglect internal field effects.}
\label{fig:DFT-p-alpha}
\end{figure}

For each value of $d$, we calculate the dipole moment as a function of electric field, $p(\Ep)$, as shown in Fig.~\ref{fig:DFT-p-alpha} (a). We see that, at smaller values electric field strength, $\Ep < 0.27 \ \text{V \AA$^{-1}$}$, changing the distance has little effect on the polarizability $\a$. As mentioned previously, we attribute this to internal field effects. In order to neglect these effects, we measure the polarizability $\a(d)$ at field strengths which are large enough to overcome the internal fields. At stronger values of field $\a(d)$ increases linearly with $d$, see Fig.~\ref{fig:DFT-p-alpha} (b).

\subsection*{Parameterization in configuration space}

\begin{figure}[h]
\centering
%\hspace*{-0.25cm}
\includegraphics[width=0.65\linewidth]{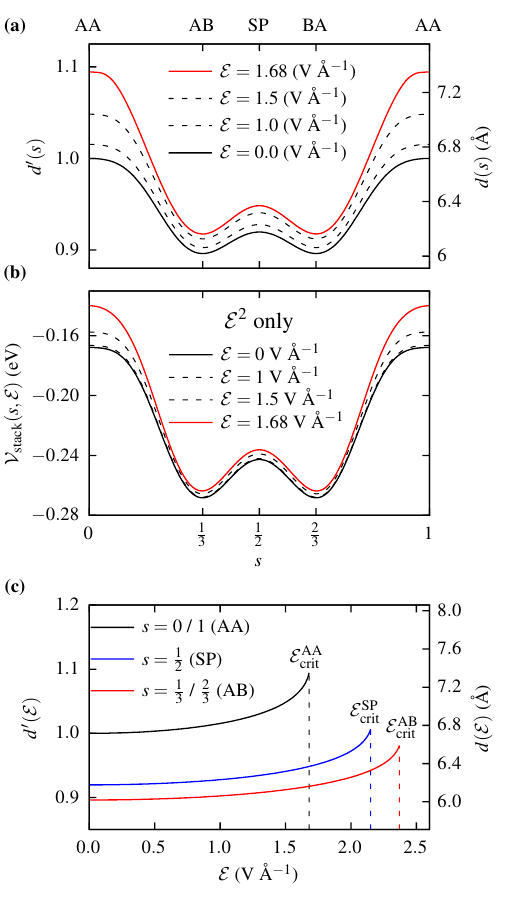}
\caption{Effect of the quadratic electrostatic term on the layer separation and stacking energy in configuration space. \textbf{(a)}: Layer separation along the configuration space diagonal for different values of $\Ep$ for 3R \chem{MoS_2}. \textbf{(b)}: Stacking energy along the diagonal in configuration space at different field strengths, including only the quadratic electrostatic term. \textbf{(c)}: Layer separation as a function of $\Ep$ for the AA, SP and AB stacking configurations. The stacking configurations have critical field values $\Ecrit^{\text{AA}} = 1.68 \ \text{V \AA$^{-1}$}$, $\Ecrit^{\text{SP}} = 2.15 \ \text{V \AA$^{-1}$}$ and $\Ecrit^{\text{AB}} = 2.37 \ \text{V \AA$^{-1}$}$, respectively.}
\label{fig:slide-d-s-E}
\end{figure}

\begin{figure}[h]
\centering
%\hspace*{-0.25cm}
\includegraphics[width=0.65\linewidth]{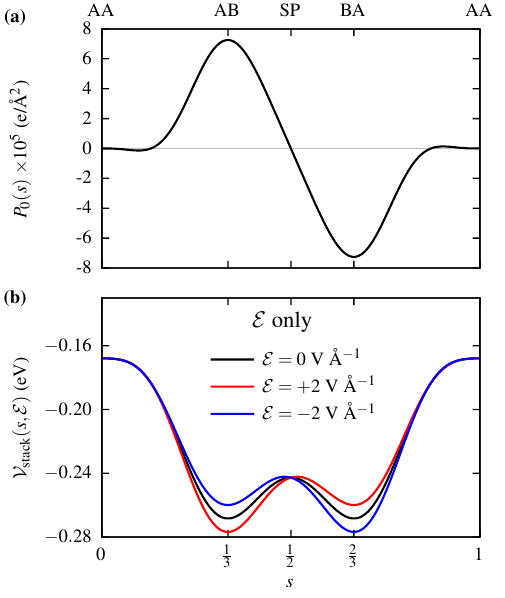}
\caption{Spontaneous polarization and its effect on the stacking energy in configuration space for 3R \chem{MoS_2}. \textbf{(a)}: Spontaneous polarization $P_0$ along the diagonal through configuration space for 3R \chem{MoS_2}. \textbf{(b)}: Stacking energy along the diagonal in configuration space at different field strengths, including only the linear electrostatic term.}
\label{fig:slide-V-s-E}
\end{figure}

Calculations were repeated to parameterize $\V_0$, $d_0$, $\a_0$ and $\a_1$ as a function of $\mathbf{s}$. The high symmetry stacking configurations AA (metal over metal), AB/BA (metal over chalcogen), and the saddle point (SP) all lie along the configuration space diagonal. Thus, it was sufficient to perform a series of calculations along the diagonal, and use a 2D Fourier interpolation to parameterize the model everywhere in configuration space. One layer was fixed, and the other layer was translated by ${\mathbf{s} = s(\mathbf{a}_1 + \mathbf{a}_2), \ s \in [0,1]}$, and the aforementioned quantities were measured on a fine grid of values of $s$.  The interpolation was done following similar approaches in previous studies\cite{jung2015origin,nam2017lattice,yu2017moire,carr2018relaxation}: each quantity is written as a Fourier expansion. The results from the first-principles calculations at different values of $\mathbf{s}$ are used to fit the Fourier coefficients, and a smooth interpolation of each quantity is obtained everywhere in configuration space. The reciprocal lattice vectors $\mathbf{G}$ of the same length are sorted into shells, and the first few shells are sufficient to obtain good parameterizations. The first-principles results and parameterization using the first three shells are shown in Fig.~4 for 3R and 2H stacked bilayer \chem{MoS_2}.

After parameterizing the model, we can examine the effect of an electric field on the system in configuration space. Fig.~\ref{fig:slide-d-s-E} (a) shows the layer separation in configuration space at different electric field values for 3R \chem{MoS_2}. Without lattice relaxation, only the quadratic electrostatic term affects the layer separation. We can see that the layer separation increases non-uniformly in configuration space, leading to a corresponding reduction in stacking energy, shown in Fig.~\ref{fig:slide-d-s-E} (b). We only show up to $\Ep = 1.68 \ \text{V \AA$^{-1}$}$ because that is the smallest critical field at which $\dmin\to\infty$ somewhere in configuration space. We examine this in more detail in Fig.~\ref{fig:slide-d-s-E} (c) by showing $\dmin$ as a function of electric field at the AA, AB and SP points. We can see that the critical field values for the AB and SP points are considerably larger. We could reach stronger field strengths by including terms proportional to $\grad d$ in the elastic energy from von Karman plate theory \cite{jung2015origin}, but this would make the model considerably more difficult to solve, requiring either solutions to fourth order differential equations or a more difficult optimization problem. After lattice relaxation, the AA regions shrink considerably, and the critical fields of interest for the domain wall are $\Ecrit^{\text{SP}}$ and $\Ecrit^{\text{AB}}$, since the domain wall is across the path ${\text{AB}\to \text{SP}\to \text{BA}}$. However, without including nonlinear terms in the elastic energy, we are limited by the smallest critical field, $\Ecrit^{\text{AA}}$.

In Fig.~\ref{fig:slide-V-s-E} (a) we show the spontaneous polarization $P_0(s)$ along the diagonal in configuration space. The coupling between the polarization and the field leads to an even splitting of the AB and BA wells, increasing one and decreasing the other, which is shown in Fig.~\ref{fig:slide-V-s-E} (b).

\clearpage

\section*{Supplementary Note 3: Frenkel-Kontorova model}

\begin{figure}[h]
\centering
%\hspace*{-0.25cm}
\includegraphics[width=0.65\columnwidth]{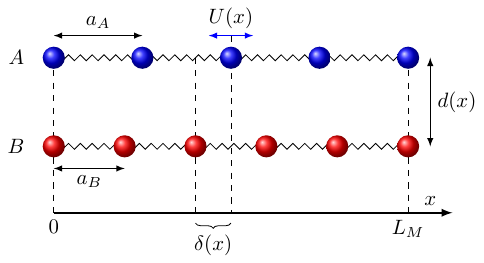}
\caption{Sketch of a 1D FK model. The layers are represented by chains of atoms A and B, connected by springs. The supercell period of the chains $L_{\text{M}}$ is shown along the $x$-axis. The lattice mismatch plays the role of the twist angle.}
\label{fig:diagram-FK}
\end{figure}

The Frenkel-Kontorova (FK) model was first introduced in the 1930s and is widely used in condensed matter physics\cite{frenkel1938frenkel,braun1998nonlinear}. It is a discrete model where a chain of atoms is subject to a rigid periodic potential, from a substrate for example\cite{frenkel1938frenkel,mcmillan1976theory,bulaevski1978commensurability,pokrovskii1978phase}. The FK model has been used as a 1D representation of twisted bilayer systems\cite{popov2011commensurate,nam2017lattice,cazeaux2017analysis}, replacing the rigid substrate with a second chain of atoms which can also deform, see Fig.~\ref{fig:diagram-FK}, and has been used to describe commensurate-incommensurate phase transitions\cite{popov2011commensurate,lebedeva2016dislocations}, lattice relaxation and domain structures\cite{nam2017lattice,carr2018relaxation,lebedeva2019energetics}.

Because the dielectric response does not break any symmetries, we study its effect on the domain walls using a 1D FK model. For 3R \chem{MoS_2}, the domain walls are along the path ${\text{AB}\to \text{SP}\to \text{BA}}$, so we use those points to parameterize the stacking energy\cite{popov2011commensurate,lebedeva2016dislocations,lebedeva2019commensurate} in a one dimensional version of configuration space,
\beq{}
\begin{split}
\V_{0}^{\text{FK}}(s) &=  \V_0^+ + \V_0^-\cos{\lb 2\pi s\rb}\\
\V_0^{\pm} &= \frac{1}{2}\lb \V_0\lb\tfrac{1}{2}\rb \pm \V_0\lb \tfrac{1}{3}\rb\rb
\end{split}
\eec
described by a single variable $s$, the relative displacement between the atoms in the two chains, and similarly for $d_0$, $\a_0$ and $\a_1$.

\begin{figure}[h]
\centering
%\hspace*{-0.25cm}
\includegraphics[width=0.85\columnwidth]{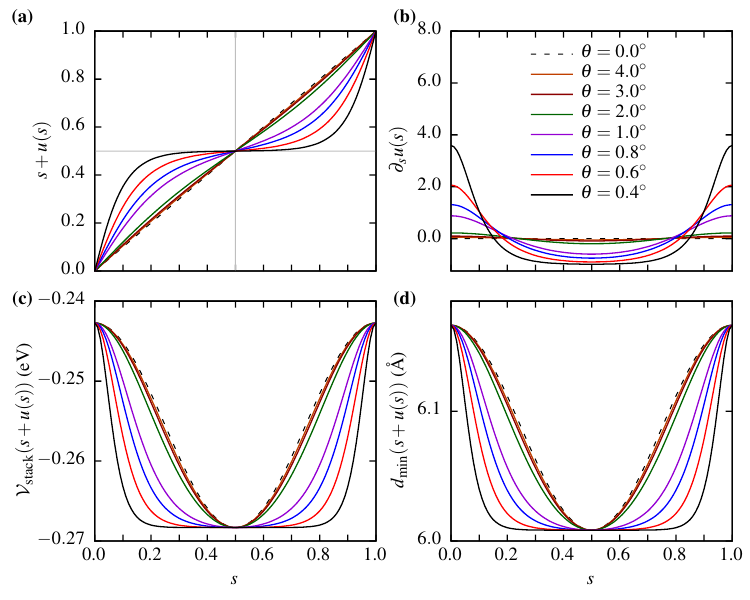}
\caption{Solutions to Eq.~\eqref{eq:ODE_FK_1} at zero field for various twist angles: \textbf{(a)}: total displacement in configuration space $s+u(s)$, \textbf{(b)}: change in displacement $\partial_su(s)$, equivalent to the strain tesnor in 1D, \textbf{(c)}: stacking potential as a function of total displacement $V_{\text{stack}}(s+u(s))$ and \textbf{(d)}: equilibrium layer separation as a function of total displacement $\dmin(s+u(s))$.}
\label{fig:ode}
\end{figure}

The total energy in configuration space is
\beq{eq:V_tot_FK}
\begin{split}
V_{\text{tot}}^{\text{FK}} = \frac{1}{\As}\int_{\As}\bigg[ \frac{1}{2}B\th^2 \lb \partial_s u\rb^2 + \V_{\text{stack}}^{\text{FK}}(s+u(s),d) + \V_{\text{elec}}^{\text{FK}}(s+u(s),d,\Ep) \bigg] \dd s
\end{split}
\eec
where $\th$ is the lattice mismatch, $B$ is the bulk modulus of \chem{MoS_2}\cite{carr2018relaxation} and $\V_{\text{stack}}$ and $\V_{\text{elec}}$ include the displacement $u(s)$ to allow for relaxations. Eq.~\ref{eq:V_tot_FK} can be also be minimized using optimization techniques, but in 1D the differential equations are easy enough to solve. Minimizing with respect to both $u$ and $d$, we get
\beq{eq:ODE_FK_1}
\begin{split}
\partial_u \bigg[\V_{\text{stack}}^{\text{FK}}(s+u(s),d) + \V_{\text{elec}}^{\text{FK}}(s+u(s),d,\Ep)\bigg] - B\th^2\partial_s^2 u(s) &= 0\\
\partial_d \bigg[\V_{\text{stack}}^{\text{FK}}(s+u(s),d) + V_{\text{elec}}^{\text{FK}}(s+u(s),d,\Ep)\bigg] &=  0
\end{split}
\eep
The second equation can be solved independently to obtain $\dmin(s,\Ep)$. This is inserted into the first equation, which can be solved numerically.

In Fig.~\ref{fig:ode} we show results for various values of $\th$ at zero field. For larger $\th$, the atoms do not move much from their initial stacking configuration. As $\th$ decreases, the elastic energy is reduced, and the atoms relax more. We see from Fig.~\ref{fig:ode} (a), the total displacement $s+u(s)$, that the atoms move to maximize the area around $s=\frac{1}{2}$, the stacking configuration with lowest energy. Figs.~\ref{fig:ode} (c) and (d) show the effect of the lattice relaxation on the stacking energy and layer separation. We can see that a domain structure forms, with wide AB/BA regions and narrow SP regions. The two are separated by a domain wall, the width of which is proportional to $\th$. Fig.~\ref{fig:ode} (b) shows the change in displacement $\partial_s u(s)$, which is proportional to the strain tensor in 1D, from which we see that there is a large strain gradient across the domain wall.

%In Fig.~3 we show results for finite electric fields at $\th = 0.4^{\circ}$. Fig.~3 (a) shows that, as the field strength increases, the atoms move back towards their initial stacking configuration. This can be understood from Figs.~3 (c) and (d): when a field is applied, the equilibrium separation increases everywhere in configuration space, which reduces the stacking energy. This reduces the atoms' ability to relax, leading to a softer domain structure.

\begin{figure}[h]
\centering
%\hspace*{-0.25cm}
\includegraphics[width=0.65\columnwidth]{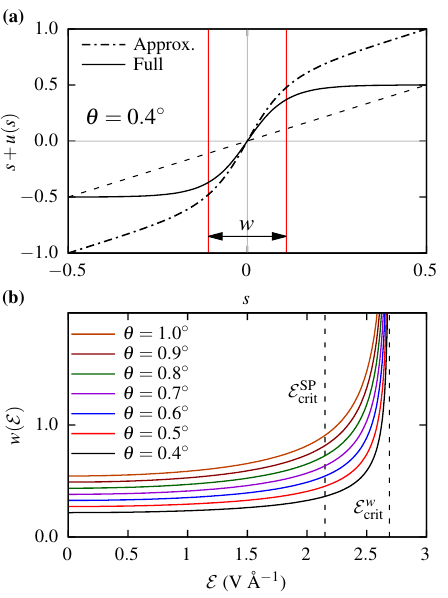}
\caption{Analysis of the domain wall width from the 1D FK model. $\textbf{(a)}$: Displacement $u(s)$ obtained by solving Eq.~\eqref{eq:ODE_FK_1} numerically and from the analytic approximation Eq.~\eqref{eq:u_approx} about $s=0$. The domain width Eq.~\eqref{eq:w_approx} is indicated by the vertical red lines. $\textbf{(b)}$: $w(\Ep)$ for various lattice mismatches from Eq.~\eqref{eq:w_approx}. The dashed lines indicate the minimum critical field in the FK model, $\Ecrit^{\text{SP}} = 2.15 \ \text{V \AA$^{-1}$}$ and the critical field at which the domain width diverges, $\Ecrit^w \approx 2.47 \ \text{V \AA$^{-1}$}$.}
\label{fig:domain-wall-efield}
\end{figure}

At $s=0$, we have the analytic approximation\cite{popov2011commensurate}:
\beq{eq:u_approx}
u(s) = \frac{2}{\pi}\tan^{-1}{\lb e^{2\pi\sqrt{\frac{V_0^{-}-\frac{1}{2}\eo\Ep^2\a_0^{-}}{B\th^2}}s}\rb} - \frac{1}{2}
\eec
which is shown alongside the numerical solutions to Eq.~\eqref{eq:ODE_FK_1} in Fig.~\ref{fig:domain-wall-efield} (a). We can approximate the width of the domain wall as
\beq{eq:w_approx}
w \sim \frac{1}{2}\sqrt{\frac{B\th^2}{\lb V_0^- -\frac{1}{2}\eo\Ep^2\a_{0}^-\rb}}
\eeq
illustrating the dependence on both twist angle and electric field. In Fig.~\ref{fig:domain-wall-efield} (b) we plot $w(\Ep)$ for several values of $\th$. $w$ diverges at a critical field $\Ecrit^w = \sqrt{\frac{2V_0^-}{\eo\a_0^-}}$ which is independent of the lattice mismatch. For \chem{MoS_2}, we have ${\Ecrit^w \approx 2.47 \ \text{V \AA$^{-1}$}}$ which is larger than the largest critical field ${\Ecrit^{\text{AB}} = 2.37 \ \text{V \AA$^{-1}$}}$.

\clearpage

\section*{Supplementary Note 4: Additional Data on twist angle and electric field dependence of stacking domains}

\begin{figure}[h!]
\centering
%\hspace*{-0.25cm}
\includegraphics[width=0.91\columnwidth]{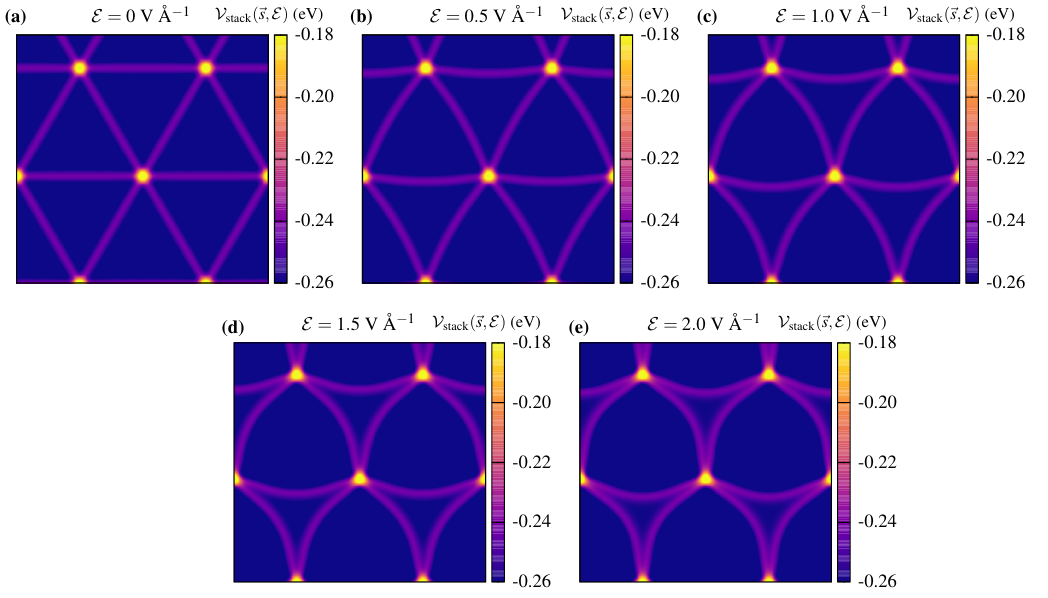}
\caption{Lattice relaxtion for 3R stacked bilayer \chem{MoS_2} at a twist angle of $\theta=0.4^{\circ}$ for several electric field values.}
\label{fig:contour04}
\end{figure}

\begin{figure}[h!]
\centering
%\hspace*{-0.25cm}
\includegraphics[width=0.91\columnwidth]{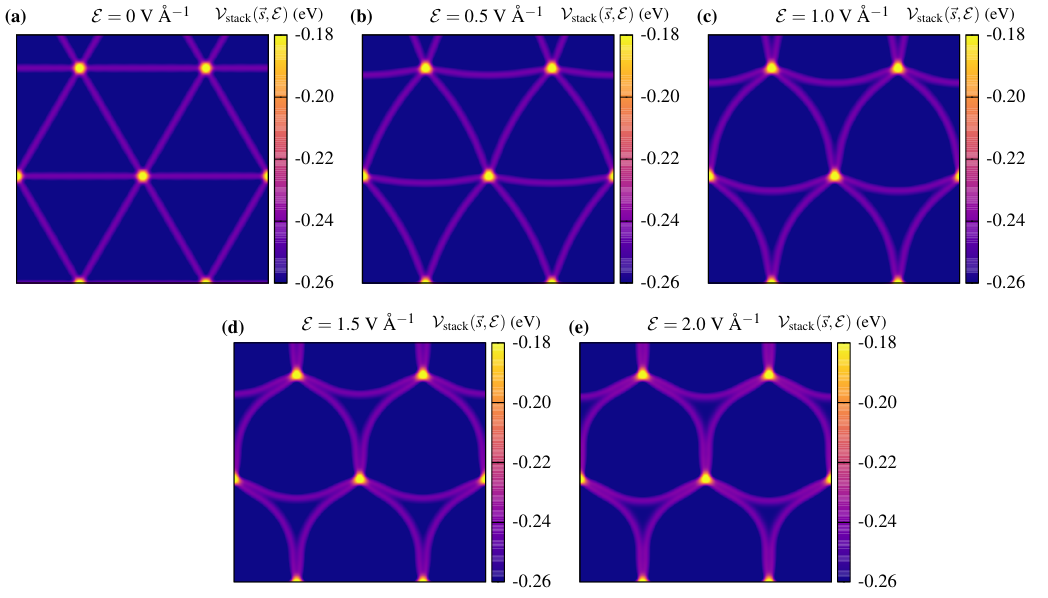}
\caption{Lattice relaxtion for 3R stacked bilayer \chem{MoS_2} at a twist angle of $\theta=0.3^{\circ}$ for several electric field values.}
\label{fig:contour03}
\end{figure}

\begin{figure}[h!]
\centering
%\hspace*{-0.25cm}
\includegraphics[width=0.91\columnwidth]{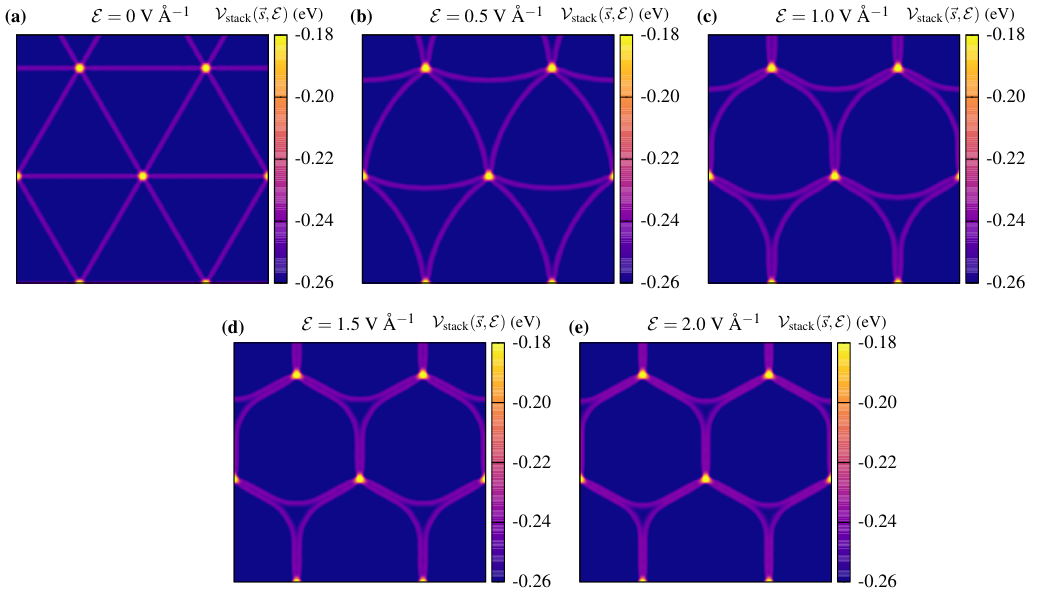}
\caption{Lattice relaxtion for 3R stacked bilayer \chem{MoS_2} at a twist angle of $\theta=0.2^{\circ}$ for several electric field values.}
\label{fig:contour02}
\end{figure}

\begin{figure}[h!]
\centering
%\hspace*{-0.25cm}
\includegraphics[width=0.91\columnwidth]{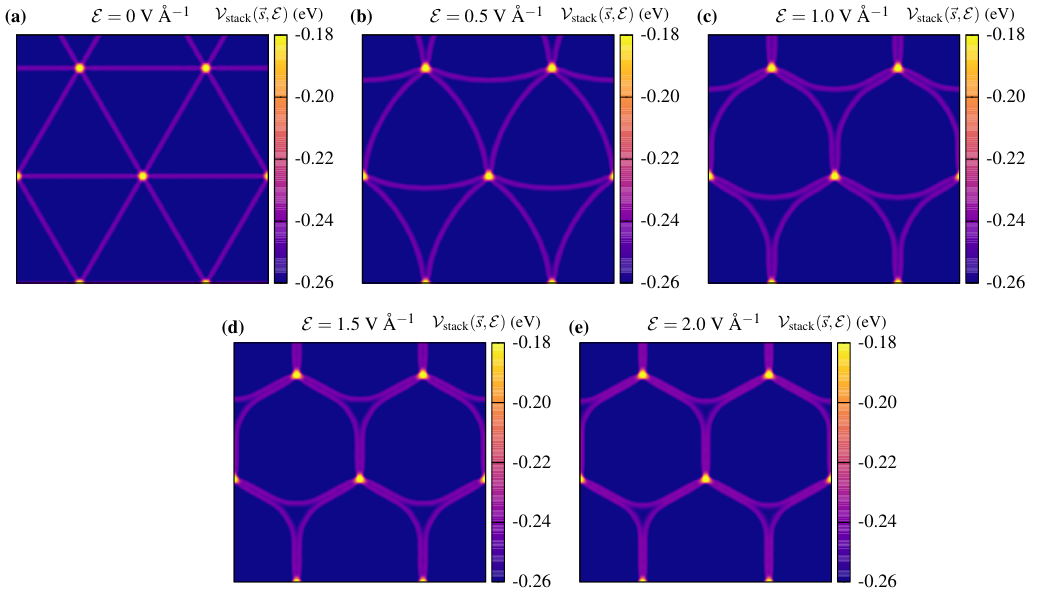}
\caption{Lattice relaxtion for 3R stacked bilayer \chem{MoS_2} at a twist angle of $\theta=0.1^{\circ}$ for several electric field values.}
\label{fig:contour01}
\end{figure}

\clearpage